\documentclass[preprint,dvipdfmx]{ptephy_v1}

\usepackage{braket}

\preprintnumber{XXXX-XXXX} 

\def\Journal#1#2#3#4{{#1} {\bf #2}, #3 (#4)}

\def\EPJC{Eur. Phys. J. C}

\def\JHEP{J. High Energy Phys.}

\def\JPG{J. Phys. G}

\def\NPB{Nucl. Phys. B}

\def\PLB{{Phys. Lett.} B}

\def\PLBOLD{Phys. Lett.}

\def\PRL{Phys. Rev. Lett.}
\def\PRD{Phys. Rev. D}

\def\RMP{Rev. Mod. Phys.}
\def\RPP{Rep. Prog. Phys.}

\begin{document}

\title{Scalar clockwork and flavor neutrino mass matrix}


\author{Teruyuki Kitabayashi}
\affil{Department of Physics, Tokai University,\\
4-1-1 Kitakaname, Hiratsuka, Kanagawa 259-1292, Japan\\
\email{teruyuki@tokai-u.jp}}


\begin{abstract}%
We study the capability of generating correct flavor neutrino mass matrix in a scalar clockwork model. First, we assume that the flavor structure is controlled by the Yukawa couplings as same as the standard model. In this case, the correct flavor neutrino mass matrix could be obtained by appropriate Yukawa couplings $Y_{\ell^\prime\ell}$ where $\ell^\prime, \ell = e, \mu, \tau$.  Next, we assume that the Yukawa couplings are extremely democratic $|Y_{\ell^\prime\ell} |=1$. In this case, the model parameters of the scalar clockwork sector, such as the site number of a clockwork gear in a clockwork chain, should have the flavor indices $\ell^\prime$ and/or $\ell$ to generate correct flavor neutrino mass matrix. We show some examples of the assignments of the flavor indices which can yield the correct flavor neutrino mass matrix.
\end{abstract}

\subjectindex{B40,B54}

\maketitle

\section{Introduction\label{sec:introduction}}
Understanding the nature of the tiny neutrino masses as well as their mixings is one of the outstanding problems in particle physics and cosmology \cite{King2015JPG}. Many theoretical mechanisms to generate tiny neutrino masses are proposed, such as seesaw mechanisms \cite{Minkowski1977PLB,Yanagida1979,Gell-Mann1979,Mohapatra1980PRL}, radiative mechanisms \cite{Zee1980PLB,Wolfenstein1980NPB,Petcov1982PLB,Zee1985PLB,Zee1986NPB,Babu1988PLB,Cheng1988PRL,Schechter1992PLB}, and the scotogenic model \cite{Ma2006PRD}. On the other hand, the neutrino mixings have been studied under assumptions of the existence of underlying flavor symmetries in the theories (for reviews, see  \cite{Altarelli2010RMP,King2013RPP,Xing2016RPP}).  Apart from the neutrino problems, there are many mysteries related to hierarchy in the particle physics.

The clockwork mechanism \cite{Giudice2017JHEP} provides a natural way to obtain the hierarchical masses and couplings in a theory. The basic idea of the clockwork mechanism is simple \cite{Teresi2017arXiv}. A product
\begin{eqnarray}
\frac{1}{q} \times \frac{1}{q} \times \cdots \times \frac{1}{q},
\end{eqnarray}
with $q > 1$, can become tiny if the number of factors increased. There is an analogy between a series of the gears in a clock and this product. In a series of the gears, large (small) movement of the gear in one side of the series can generate a small (large) movement of the gear in the opposite side. The factor $1/q$ behaves like a clockwork gear and the product behaves like a series of the gears.

To implement this idea in quantum field theory, a large number of fields $\phi_i$ are introduced as the clockwork gears to a theory. These fields interact with the standard model (SM) particles schematically as
\begin{eqnarray}
\phi_0 \underset{\frac{1}{q}}{-} \phi_1 \underset{\frac{1}{q}}{-} \cdots \underset{\frac{1}{q}}{-} \phi_N - {\rm SM},
\end{eqnarray}
with couplings $1/q \lesssim 1$, where $N$ denotes the number of gears. The series of the fields behaves like a clockwork chain. If one of the mass eigenstate (typically the lightest state) $\phi_{\rm light}$ is essentially given by $\phi_0$, the interaction between $\phi_{\rm light}$ and the standard model particles will be suppressed as
\begin{eqnarray}
\phi_{\rm light} - {\rm SM} \sim \frac{1}{q^N},
\end{eqnarray}
for large $N$. Therefore, we can obtain a tiny coupling $1/X$ by $\mathcal{O}(1)$ couplings $1/q$ and a large number of fields $N \sim \log_q X$. This is the outline of the scalar clockwork mechanism. The basics of other clockwork mechanisms, such as the fermion clockwork mechanism, is essentially same as the basics of the scalar clockwork mechanism.

The applications of the clockwork mechanism have been extensively studied in the literature, e.g., for the axion \cite{Choi2014PRD,Choi2016JHEP,Kaplan2016PRD,Farina2017JHEP,Coy2017JHEP,Agrawal2018JHEP1,Long2018JHEP,Agrawal2018JHEP2,Bonnefoy2019EPJC,Bae2019PRD}, for inflation \cite{Kehagias2017PLB,Park2018arXiv}, for dark matter \cite{Marzola2018PRD,Hambey2017JHEP,Kim2018PRD1,Goudelis2018JHEP,Kim2018PRD2}, for the muon $g-2$ \cite{Hong2018PRD}, for string theory \cite{Ibanez2018JHEP,Antoniadis2018EPJC,Im2019JHEP}, for gravity \cite{Kehagias2018JHEP,Niedermann2018PRD}, for GUTs \cite{Gersdorff2020arXiv,Babu2020ArXiv} for charged fermion masses and mixings \cite{Patel2017PRD}, for quark masses and mixings \cite{Alonso2018JHEP} and for Goldstone bosons \cite{Ahmed2017PRD}. 

The applications of the clockwork mechanism for the neutrino sector have been studied for tiny neutrino masses \cite{Park2018PLB,Banerjee2018JHEP,Hong2019JHEP} and for their mixings \cite{Ibarra2018PLB, Kitabayashi2019PRD}. Up to now, there are two fermion clockwork models for the neutrino mixings \cite{Ibarra2018PLB, Kitabayashi2019PRD}; however, there is no scalar clockwork model for the neutrino mixings.

In this paper, towards a construction of the scalar clockwork models including neutrino mixings, we extend the scalar clockwork model proposed by Banerjee, Ghosh and Ray \cite{Banerjee2018JHEP} for one generation neutrino (without mixing) to a model for three generation neutrinos (with mixings). Since any correct scalar clockwork models for three generation neutrinos should yield the $3 \times 3$ flavor neutrino mass matrix which is consistent with observations, we would like to concentrate our discussion on the mathematical capability of generating correct flavor neutrino mass matrix.

The paper is organized as follows. In Sec.\ref{section:review}, we present a brief review of the scalar clockwork mechanisms.  In Sec.\ref{section:mass_matrix}, towards a construction of the scalar clockwork models including neutrino mixings, we study the mathematical capability of generating correct flavor neutrino mass matrix in a scalar clockwork model.  Section \ref{section:summary} is devoted to a summary.

\section{Review of scalar clockwork \label{section:review}}
\subsection{Scalar clockwork mechanism}
The total Lagrangian of the standard model with the clockwork sector reads
\begin{eqnarray}
\mathcal{L} =\mathcal{L}_{\rm SM} + \mathcal{L}_{\rm CW} + \mathcal{L}_{\rm SM-CW},
\end{eqnarray}
where $\mathcal{L}_{\rm SM}$ denotes the standard model Lagrangian, $\mathcal{L}_{\rm CW}$ denotes the interactions in the clockwork sector and $\mathcal{L}_{\rm SM-CW}$ denotes the interactions between the standard model sector and the clockwork sector. 

In the scalar clockwork models, there are $N+1$ scalars, $\Phi_j$ ($j=0,1,\cdots,N$) with $N+1$ global U(1) symmetries. These U(1) symmetries are spontaneously broken to their discrete subgroups ${\rm Z}_2$ at some scale $f$. The clockwork Lagrangian can be written as \cite{Giudice2017JHEP,Banerjee2018JHEP}
\begin{eqnarray}
\mathcal{L}_{\rm CW} = \sum_{j=0}^N \left[ \partial_\mu \Phi_j^\dag \partial^\mu \Phi_j -\frac{\lambda}{8}(\Phi_j^\dag \Phi_j-f^2)^2 \right]  
 + \frac{1}{2}\Lambda^{3-q}\sum_{j=0}^{N-1} \left(\Phi_j^\dag \Phi_{j+1}^q + {\rm h.c.}\right),
\label{Eq:Lcw}
\end{eqnarray}
where $q \in  \mathbb{Z}$ as well as $j \in \mathbb{N}$ \cite{Ahmed2017PRD}. The first two terms in Eq.(\ref{Eq:Lcw}) are invariant under the global ${\rm U}(1)^{N+1}$, on the other hand, the last term breaks the symmetry down to a single remnant ${\rm U}(1)_{\rm CW}$.  The explicit breaking term is renormalizable and soft ($\Lambda \ll f$) if $1 < q \le 3$ is satisfied \cite{Kaplan2016PRD,Banerjee2018JHEP}. Since $q \in  \mathbb{Z}$, the requirement of
\begin{eqnarray}
q=2,3,
\label{Eq:q23}
\end{eqnarray}
should be satisfied in the scalar clockwork models described by the Lagrangian in Eq.(\ref{Eq:Lcw}). Sometimes, the requirements of $q \in \mathbb{N}$ as well as $j \in \mathbb{N}$ are relaxed in the analysis (see, for examples, Ref.\cite{Kim2018PRD1}); however, we would like to keep the requirements of Eq.(\ref{Eq:q23}) and $j \in \mathbb{N}$ in the main part of this paper.

After the spontaneous symmetry breaking, the effective fields are $N+1$  Nambu-Goldstone (NG) bosons $\pi_j$ that can be conveniently described
\begin{eqnarray}
U_j = e^{i\pi_j/f}.
\end{eqnarray}
The explicit breaking term in $\mathcal{L}_{\rm CW}$ is not invariant under the shift symmetries of the NG bosons ($\pi_j \rightarrow \pi_j + \alpha_j$). On the other hand, the remnant unbroken ${\rm U}(1)_{\rm CW}$ is invariant under the transformation
\begin{eqnarray}
\pi_j \rightarrow \pi_j + \frac{\alpha}{q^{j-1}}, \quad j=0,\cdots,N.
\end{eqnarray}
The unbroken ${\rm U}(1)_{\rm CW}$ corresponds to the generator
\begin{eqnarray}
Q = \sum_{j=0}^N \frac{Q_j}{q^j},
\end{eqnarray}
where $Q_j$ is the generator of j-th site. 

In terms of the fields $\pi_j$, we obtain the pseudo-NG boson potential \cite{Banerjee2018JHEP}
\begin{eqnarray}
V_\pi &=& -\frac{1}{2}f^{q-1} \Lambda^{3-q} \sum_{j=0}^{N-1} \left(U_j^\dag U_{j+1}^q + {\rm h.c.}\right) \nonumber \\
&=& -f^{q-1}\Lambda^{3-q}\sum_{j=0}^{N-1} \cos \left( \frac{\pi_j - q\pi_{j+1}}{f} \right) \nonumber \\
&=& -\frac{1}{2}\sum_{i,j=0}^{N} \pi_i (M_{\pi}^2)_{ij} \pi_j + \mathcal{O}(\pi^4) .
\end{eqnarray}
The mass matrix is given by
\begin{eqnarray}
M^2_\pi &=& f^{q-1}\Lambda^{3-q} \left( 
\begin{array}{cccccc}
1 & -q  & 0 & \cdots & 0 & 0\\
-q & q^2+1  & -q & \cdots & 0 & 0\\
0 & -q  & q^2+1 & \cdots & 0 & 0\\
\vdots & \vdots  & \vdots & \ddots & \vdots & \vdots\\
0& 0& 0 & & q^2+1 & -q \\
0& 0& 0 & \cdots & -q & q^2 \\
\end{array}
\right). 
\end{eqnarray}
The tridiagonal symmetric mass matrix $M_\pi^2$ can be diagonal by an orthogonal rotation
\begin{eqnarray}
\pi_j = \mathcal{O}_{jk}a_k, \quad (j=0,\cdots,N, \quad k=1, \cdots, N),
\end{eqnarray}
where 
\begin{eqnarray}
\mathcal{O}_{j0} &=& \frac{1}{q^j}\sqrt{\frac{q^2-1}{q^2-q^{-2N}}},  \label{Eq:Ojk} \\
\mathcal{O}_{jk} &=& \sqrt{\frac{2}{(N+1)\lambda_k}}\left[ q\sin\frac{jk\pi}{N+1}-q\sin\frac{(j+1)k\pi}{N+1} \right], \nonumber
\end{eqnarray}
with
\begin{eqnarray}
\lambda_k = q^2+1-2q\cos\frac{k\pi}{N+1}.
\end{eqnarray}
After the rotation, one massless eigenvalue of the one massless NG mode $a_0$ and $N$ massive eigenvalues for $N$ massive pseudo-NG modes $a_k$ are obtained as
\begin{eqnarray}
m_{a_0}^2=0, \quad m_{a_k}^2 = \lambda_k f^{q-1} \Lambda^{3-q}, \quad (k=1,\cdots,N).
\end{eqnarray}

$\mathcal{O}_{j0}$ in Eq.(\ref{Eq:Ojk}) measures the component of the massless NG state contained in $\pi_j$. Since $\mathcal{O}_{j0} \propto q^{-j}$, the NG state $a_0=\mathcal{O}_{j0}\pi_j$ is $q$ times smaller than for the previous site. Thus, the NG interaction may be secluded away from the last side for large $N$. If a standard model fields are coupled to the clockwork sector only through its $N$-th site, the massless eigenstate $a_0$ is hierarchically localized at the different sites with a factor $1/q^j$ and can give rise to an exponential suppression. This is the scalar clockwork mechanism.

\subsection{Scalar clockwork and one flavor neutrino}
A way to generate the tiny neutrino mass by the scalar clockwork mechanism for one flavor neutrino is proposed by Banerjee, Ghosh and Ray \cite{Banerjee2018JHEP}. They apply {\it clockworked vacuum expectation values (VEVs)} mechanism to a simple model to explain the tiny neutrino mass. 

The NG bosons arising in $V_\pi$ posses a discrete ${\rm Z_2}$ symmetry and can not receive VEVs. In the clockworked VEVs mechanism, to generate a hierarchical VEV structure, an additional soft breaking potential ($\mu_1, \mu_2 \ll f$)
\begin{eqnarray}
V_{\rm soft} &=& -\frac{\mu_1^2 f^2}{4} (U_k+{\rm h.c.} )^2 + \frac{\mu_2^3f}{2}(iU_k+{\rm h.c.} ) \nonumber \\
&=& -\mu_1^2 f^2 \cos^2 \frac{\pi_k}{f} - \mu_2^3 f \sin\frac{\pi_k}{f},
\end{eqnarray}
is introduced for $k$-th site to break the residual ${\rm U}(1)_{\rm CW}$ as well as ${\rm Z_2}$ symmetry explicitly. If the breaking potential added at zeroth site, $k=0$, the minimization condition for the total potential $V=V_\pi + V_{\rm soft}$ yields 
\begin{eqnarray}
\braket{\pi_0} = \frac{\mu_2^3}{2\mu_1^2},  
\end{eqnarray}
and 
\begin{eqnarray}
\braket{\pi_1} = \frac{\braket{\pi_0}}{q}, \ \braket{\pi_2} = \frac{\braket{\pi_0}}{q^2}, \cdots, \braket{\pi_N} = \frac{\braket{\pi_0}}{q^N}.
\end{eqnarray}
The VEV arising at the farthest end ($N$-th site) from the soft-breaking site ($0$-th site) may be small for large $q^N$. This is the clockworked VEVs mechanism.

A simple model for one generation neutrino can be obtained as follows. According to Banerjee et al, we assume the right-handed neutrino $\nu_R$ possesses a charge under the $Z_2$ symmetry of the $j$-th site of the clockwork chain, denoted by $Z_2^{(j)}$. The $Z_2^{(j)}$ charges are assigned as $Z_2^{(j)}(\pi_j)=Z_2^{(j)}(\nu_R)=-1$ and $Z_2^{(j)}({\rm others})=+1$. In this case, the interaction between the clockwork sector and the right-handed neutrino will be schematically
\begin{eqnarray}
\pi_0 - \pi_1 - \pi_2 -\cdots -  &\pi_j &  - \cdots -   \pi_N.  \nonumber \\
&|& \\
&\nu_R& \nonumber
\end{eqnarray}
This phenomenon is described by the following interaction Lagrangian
\begin{eqnarray}
\mathcal{L}_{\rm SM-CW}= y  \left( \frac{\pi_j}{f} \right)  \bar{\ell}_L \tilde{H} \nu_{R} + {\rm h.c.},
\label{Eq:LoneNu}
\end{eqnarray}
where $y$ denotes some effective coupling, $\ell_L$ denotes the standard model left-handed lepton doublet, and $H$ denotes the standard model Higgs doublet. After symmetry breaking, the fields obtain VEV:
\begin{eqnarray}
&&\braket{\pi_0}- \frac{\braket{\pi_0}}{q}-\frac{\braket{\pi_0}}{q^2} - \cdots -  \frac{\braket{\pi_0}}{q^j}   - \cdots -   \frac{\braket{\pi_0}}{q^N}, \nonumber \\
&& \hspace{45mm} | \\
&& \hspace{43mm} \nu_R \nonumber
\end{eqnarray}
and the Dirac mass of the neutrino is obtained as
\begin{eqnarray}
m_\nu =\frac{yv}{\sqrt{2}} \frac{\braket{\pi_j}}{f}  \simeq  \frac{yv}{\sqrt{2}} \left( \frac{\braket{\pi_0}}{f}\frac{1}{q^j} \right) = \frac{yv^{\rm eff}}{\sqrt{2}},
\end{eqnarray}
where $v$ denotes the VEV of the Higgs and $v^{\rm eff}$ denotes an effective VEV. The effective VEV
\begin{eqnarray}
v^{\rm eff}=v  \left( \frac{\braket{\pi_0}}{f}\frac{1}{q^j} \right),
\label{Eq:veff}
\end{eqnarray}
may be tiny for large $q^j$ by the clockworked VEVs mechanism and the tiny neutrino mass may be generated. For example, assuming 
\begin{eqnarray}
y \sim \mathcal{O}(1), \quad \frac{\braket{\pi_0}}{f} \sim \mathcal{O}(0.1),  \quad q=3,
\end{eqnarray}
we find the tiny neutrino mass $m_\nu \sim 0.1$ eV, if the right-handed neutrino couples to the $24$-th site ($j=24$) of the clockwork chain.

\section{Flavor neutrino mass matrix \label{section:mass_matrix}}
\subsection{Experimental constraints}
We show the basics of the flavor neutrino mass matrix and the constraints on the mass matrix form the observations. 

The flavor neutrino mass matrix 
\begin{eqnarray}
M = \left( 
\begin{array}{ccc}
M_{ee} & M_{e\mu}  & M_{e\tau}  \\
M_{\mu e} & M_{\mu \mu}  & M_{\mu \tau}  \\
M_{\tau e} & M_{\tau\mu}  & M_{\tau\tau}  \\
\end{array}
\right) ,
\end{eqnarray}
satisfies the relation
\begin{eqnarray}
M M^\dag = U_{\rm PMNS}  \left( 
\begin{array}{ccc}
m_1^2 & 0 & 0 \\
0 & m_2^2  & 0 \\
0 & 0  & m_3^2 \\
\end{array}
\right) U_{\rm PMNS}^\dag,
\end{eqnarray}
where $m_1,m_2$ and $m_3$ denote the neutrino mass eigenstates and 
\begin{eqnarray}
U_{\rm PMNS}=  
\ \left( {\begin{array}{*{20}{c}}
c_{12}c_{13} & s_{12}c_{13} & s_{13}\\
- s_{12}c_{23} - c_{12}s_{23}s_{13} & c_{12}c_{23} - s_{12}s_{23}s_{13} & s_{23}c_{13}\\
s_{12}s_{23} - c_{12}c_{23}s_{13} & - c_{12}s_{23} - s_{12}c_{23}s_{13} & c_{23}c_{13}
\end{array}} \right), \nonumber 
\label{Eq:U_PDG}
\end{eqnarray}
denotes the mixing matrix \cite{PDG}. We use the abbreviations $c_{ij}=\cos\theta_{ij}$ and $s_{ij}=\sin\theta_{ij}$  ($i,j$=1,2,3) and ignore the $CP$-violating phase. 

Although the neutrino mass ordering (either the normal mass ordering or the inverted mass ordering) is not determined, a global analysis shows that the preference for the normal mass ordering is mostly due to neutrino oscillation measurements \cite{Salas2018PLB}. Upcoming experiments for neutrinos will be solve this problem \cite{Aartsen2020PRD}. In this paper, we assume the normal mass hierarchical spectrum for the neutrinos, e.g., $m_1<m_2<m_3$.  The best-fit values of the squared mass differences $\Delta m_{ij}^2=m_i^2-m_j^2$ and the mixing angles (as well as $1 \sigma$ and $3 \sigma$ allowed regions) are estimated as \cite{Esteban2019JHEP}
\begin{eqnarray} 
\frac{\Delta m^2_{21}}{10^{-5} {\rm eV}^2} &=& 7.39^{+0.21}_{-0.20} \quad (6.79\rightarrow 8.01), \nonumber \\
\frac{\Delta m^2_{31}}{10^{-3}{\rm eV}^2} &=& 2.528^{+0.029}_{-0.031}\quad (2.436 \rightarrow 2.618), \nonumber \\
\theta_{12}/^\circ &=& 33.82^{+0.78}_{-0.76} \quad (31.61 \rightarrow 36.27), \nonumber \\
\theta_{23}/^\circ &=& 48.6^{+1.0}_{-1.4} \quad (41.1 \rightarrow 51.3), \nonumber \\
\theta_{13}/^\circ &=& 8.60^{+0.13}_{-0.13}\quad (8.22 \rightarrow 8.98),
\label{Eq:neutrino_observation}
\end{eqnarray}
where the $\pm$ denote the $1 \sigma$ region and the parentheses denote the $3 \sigma$ region. 

In this paper, we will use the following experimental constraints on the flavor neutrino mass matrix.

\

\begin{figure}[t]
\begin{center}
\includegraphics{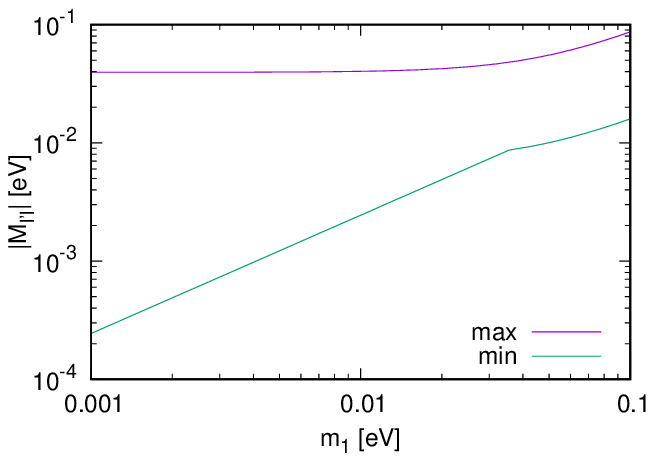}
\caption{Allowed region of the flavor neutrino masses $\left\vert M_{\ell^\prime \ell}\right\vert$ ($\ell^\prime,\ell = e,\mu, \tau$) in the $3\sigma$ region. The upper curve and the lower curve show the maximum and minimum magnitude of the flavor neutrino masses, respectively. The allowed region becomes wide (narrow) for small (large) $m_1$.}
\label{fig:MllMinMax}
\end{center}
\end{figure}

\textbf{(A) Best-fit values:} 
The flavor neutrino mass matrix should be
\begin{eqnarray}
M =  \left( 
\begin{array}{ccc}
0.821m_1 & 0.550m_2 & 0.150m_3 \\
-0.461m_1 & 0.487m_2  & 0.742m_3 \\
0.335m_1 & -0.678m_2  & 0.654m_3 \\
\end{array}
\right),
\end{eqnarray}
for the best-fit values of neutrino oscillation parameters, where
\begin{eqnarray}
m_2 = \sqrt{7.39\times 10^{-5} + m_1^2} \ {\rm eV}, \quad
m_3 = \sqrt{2.528\times 10^{-3}+ m_1^2} \ {\rm eV},
\end{eqnarray}
for $m_1$ [eV]. We will use 
\begin{eqnarray}
M=  \left( 
\begin{array}{ccc}
0.0821 & 0.0552 & 0.0617 \\
-0.0461 & 0.0489  & 0.0831 \\
0.0335 & -0.0681  & 0.0732 \\
\end{array}
\right)  {\rm eV},
\label{Eq:Mbestfit}
\end{eqnarray}
for $m_1=0.1$ eV as a benchmark of the correct flavor neutrino mass matrix with the best-fit values of neutrino parameters.

\

\textbf{(B) $3 \sigma$ region:} 
Figure \ref{fig:MllMinMax} shows the allowed region of the magnitude of the flavor neutrino masses $\left\vert M_{\ell^\prime \ell}\right\vert$ ($\ell^\prime,\ell = e,\mu, \tau$) in the $3\sigma$ region. The upper curve and the lower curve show the maximum and minimum magnitude of the flavor neutrino masses, respectively. The allowed region becomes wide for small $m_1$ and becomes narrow for large $m_1$. We will use
\begin{eqnarray}
 \left\vert M_{\ell^\prime \ell}\right\vert
 =
\begin{cases}
0.000244 - 0.0395 \ {\rm eV} &  (m_1 = 0.001\ {\rm eV}), \\
0.00244 - 0.0403 \ {\rm eV} &  (m_1 = 0.01\ {\rm eV}), \\
0.0159 - 0.0868 \ {\rm eV} &  (m_1 = 0.1\ {\rm eV}), 
\end{cases}
\label{eq:MllMaxMin}
\end{eqnarray}
as well as
\begin{eqnarray}
&&\left( 
\begin{array}{ccc}
\left\vert M_{ee} \right\vert & \left\vert M_{e\mu} \right\vert & \left\vert M_{e\tau} \right\vert \\
\left\vert M_{\mu e} \right\vert & \left\vert M_{\mu\mu} \right\vert & \left\vert M_{\mu\tau} \right\vert \\
\left\vert M_{\tau e} \right\vert & \left\vert M_{\tau\mu} \right\vert & \left\vert M_{\tau\tau} \right\vert
\end{array}
\right) \nonumber \\
&&= \left( 
\begin{array}{ccc}
0.000796 - 0.000843 & 0.00430 - 0.00527 & 0.00706 - 0.00799 \\
0.000423 - 0.000529 & 0.00359 - 0.00534 & 0.0321 - 0.0395\\
0.000244 - 0.000390 & 0.00493 - 0.00645  & 0.0305 - 0.0382
\end{array}
\right) \ {\rm eV},
\label{Eq:3sigma_0.001}
\end{eqnarray}
for $m_1=0.001$ eV,
\begin{eqnarray}
&&\left( 
\begin{array}{ccc}
\left\vert M_{ee} \right\vert & \left\vert M_{e\mu} \right\vert & \left\vert M_{e\tau} \right\vert \\
\left\vert M_{\mu e} \right\vert & \left\vert M_{\mu\mu} \right\vert & \left\vert M_{\mu\tau} \right\vert \\
\left\vert M_{\tau e} \right\vert & \left\vert M_{\tau\mu} \right\vert & \left\vert M_{\tau\tau} \right\vert
\end{array}
\right) \nonumber \\
&&= \left( 
\begin{array}{ccc}
0.00796 - 0.00843 & 0.00671 - 0.00786  &  0.00720-  0.00814 \\
0.00423 - 0.00529 &  0.00560- 0.00795  & 0.0327 - 0.0403 \\
0.00244- 0.00390 &  0.00769- 0.00961  & 0.0311 - 0.0389
\end{array}
\right) \ {\rm eV},
\label{Eq:3sigma_0.01}
\end{eqnarray}
for $m_1=0.01$ eV and
\begin{eqnarray}
&&\left( 
\begin{array}{ccc}
\left\vert M_{ee} \right\vert & \left\vert M_{e\mu} \right\vert & \left\vert M_{e\tau} \right\vert \\
\left\vert M_{\mu e} \right\vert & \left\vert M_{\mu\mu} \right\vert & \left\vert M_{\mu\tau} \right\vert \\
\left\vert M_{\tau e} \right\vert & \left\vert M_{\tau\mu} \right\vert & \left\vert M_{\tau\tau} \right\vert
\end{array}
\right)\nonumber \\
&&= \left( 
\begin{array}{ccc}
 0.0796-  0.0843&  0.0519-  0.0588 &  0.0159-  0.0175 \\
0.0423 -  0.0529&  0.0433-0.0595   &  0.0724-0.0868  \\
 0.0244- 0.0390 & 0.0596 - 0.0719  &  0.0689- 0.0838
\end{array}
\right) \ {\rm eV},
\label{Eq:3sigma_0.1}
\end{eqnarray}
for $m_1=0.1$ eV as the benchmarks of the correct flavor neutrino masses in the $3\sigma$ region.

\subsection{Yukawa dominant}
As we addressed in the previous section, the tiny neutrino mass can be generated by clockworked VEVs mechanisms without neutrino mixing. 

Now, we extend the clockworked VEVs model for one generation neutrino (without mixing) to a model for three generation neutrinos (with mixings). As we mentioned in Introduction, since any correct scalar clockwork models for three generation neutrinos should yield the $3 \times 3$ neutrino flavor mass matrix which is consistent with observations, we would like to concentrate our discussion on the mathematical capability of generating correct flavor neutrino mass matrix.

To reproduce the mixings between three Dirac neutrino flavors, the non-diagonal Yukawa matrix elements $Y_{\ell^\prime\ell}$ $(\ell^\prime, \ell=e,\nu,\tau)$ should be included to the model. As the simplest extension of the Eq.(\ref{Eq:LoneNu}), we just change one right-handed neutrino $\nu_R$ to three right-handed neutrinos $\nu_{\ell R}$ $(\ell=e,\nu,\tau)$ and the single Yukawa coupling $y$ to the nine Yukawa couplings $Y_{\ell^\prime \ell}$. The extended model has the following interaction Lagrangian
\begin{eqnarray}
\mathcal{L}_{\rm SM-CW}=\sum_{\ell^\prime, \ell} Y_{\ell^\prime\ell}  \left( \frac{\pi_j}{f} \right)  \bar{\ell}^\prime_L \tilde{H} \nu_{\ell R} + {\rm h.c.}.
\end{eqnarray}
The tiny elements of the flavor neutrino mass matrix
\begin{eqnarray}
M_{\ell^\prime \ell} \simeq Y_{\ell^\prime\ell} \frac{v}{\sqrt{2}} \left( \frac{\braket{\pi_0}}{f}\frac{1}{q^j} \right) = Y_{\ell^\prime\ell}  \frac{v^{\rm eff}}{\sqrt{2}},
\label{Eq:YukawaDominantVeff}
\end{eqnarray}
are obtained by the clockworked VEVs mechanism where the effective VEV, $v^{\rm eff}$, is the same as Eq.(\ref{Eq:veff}). 

The correct neutrino masses and mixings are obtained by an appropriate Yukawa matrix as same as the standard model. For example, the Yukawa matrix 
\begin{eqnarray}
Y=
\left( 
\begin{array}{ccc}
1.33 & 0.896 & 0.272 \\
-0.748 & 0.793  & 1.35 \\
0.544 & -1.10  & 1.19 \\
\end{array}
\right),
\end{eqnarray}
and
\begin{eqnarray}
\frac{\braket{\pi_0}}{f}=0.1, \quad 
q=3,  \quad
j=24,
\end{eqnarray}
yield the flavor neutrino mass matrix in Eq.(\ref{Eq:Mbestfit}) which is consistent with the best-fit values of neutrino oscillation parameters for $m_1=0.1$. 

We would like to point out that we can rewire the Eq.(\ref{Eq:YukawaDominantVeff}) as
\begin{eqnarray}
M_{\ell^\prime \ell} \simeq \frac{v}{\sqrt{2}} Y^{\rm eff}_{\ell^\prime\ell},
\end{eqnarray}
where 
\begin{eqnarray}
Y^{\rm eff}_{\ell^\prime\ell} = Y_{\ell^\prime\ell} \left( \frac{\braket{\pi_0}}{f}\frac{1}{q^j} \right),
\end{eqnarray}
behaves like an {\it effective  clockworked Yukawa couplings}. We can use the clockworked VEVs mechanism to realize the clockworked Yukawa couplings as well as clockworked VEVs.

Because all flavor indices are assigned to the Yukawa couplings, the structure of the flavor mixings is controlled by the Yukawa couplings $Y_{\ell^\prime\ell}$. The clockwork part $\left( \frac{\braket{\pi_0}}{f}\frac{1}{q^j} \right)$ cannot contribute to the details of the flavor structure. The clockwork part just guarantees the generation of the tiny neutrino masses even if the magnitudes of the Yukawa couplings are order one.

\subsection{Clockwork dominant\label{section:mass_matrix_cw_dominant}}
As the opposite case of the Yukawa dominant case, if we assume that the Yukawa couplings are extremely democratic \cite{Gersdorff2017JHEP}
\begin{eqnarray}
| Y_{\ell^\prime\ell}| = 1,
\end{eqnarray}
the details of the flavor structure should be controlled by the clockwork part $\left( \frac{\braket{\pi_0}}{f}\frac{1}{q^j} \right)$. In this case, the clockwork part should have the flavor indices $\ell^\prime, \ell = e, \mu, \tau$. 

Without  discussions of the physical possibility of the model building, there are several possible combinations of the assignment of the flavor indices in the clockwork part. For example, there are four possible combinations of the flavor indices $\ell^\prime$ and $\ell$ for $\pi_0$, e.g., $\pi_0^{(\ell^\prime\ell)}$, $\pi_0^{(\ell^\prime)}$, $\pi_0^{(\ell)}$ and $\pi_0$.  As same as  $\pi_0$, other three parameters, $f$, $q$ and $j$ in the clockwork part could be flavored parameters. The total number of combinations of assignment of the flavor indices is $4^4=256$. The minimum assignment of the flavor indices yields the following flavor neutrino masses
\begin{eqnarray}
\left\vert M_{\ell^\prime \ell}\right\vert = \frac{v}{\sqrt{2}} \frac{\braket{\pi_0}}{f}\frac{1}{q^j}, 
\end{eqnarray}
which is essentially same as the one flavor neutrino (without mixing) clockworked VEVs case in the previous section. On the other hand, the maximal assignment of the flavor indices yields 
\begin{eqnarray}
\left\vert M_{\ell^\prime \ell}\right\vert = \frac{v}{\sqrt{2}} \frac{\braket{\pi_0^{(\ell^\prime\ell)}}}{f_{\ell^\prime\ell}}\frac{1}{q_{\ell^\prime\ell}^{j_{\ell^\prime \ell}}}.
\end{eqnarray}

Since the  conditions $q=2,3$ and $j \in \mathbb{N}$ should be satisfied in the one flavor neutrino clockwork models, we require the following condition 
\begin{eqnarray}
q_{\ell^\prime\ell}=2,3, \quad j_{\ell^\prime\ell} \in \mathbb{N},
\end{eqnarray}
to the three neutrino flavor models. With these requirements, we have the constraints on the site number as
\begin{eqnarray}
j_{\ell^\prime \ell} &=&
\begin{cases}
36, 37, \cdots, 49 &  (q_{\ell^\prime \ell}=2) \\
23, 24,\cdots,  31&  (q_{\ell^\prime \ell}=3)
\end{cases}
\quad (m_1 = 0.001\ {\rm eV}),
\nonumber \\
j_{\ell^\prime \ell} &=&
\begin{cases}
36, 37, \cdots, 46 &  (q_{\ell^\prime \ell}=2) \\
23,24,\cdots, 29 &  (q_{\ell^\prime \ell}=3)
\end{cases}
\quad (m_1 = 0.01\ {\rm eV}),
\nonumber \\
j_{\ell^\prime \ell} &=&
\begin{cases}
35,36, \cdots, 43 &  (q_{\ell^\prime \ell}=2) \\
22,23, \cdots, 27 &  (q_{\ell^\prime \ell}=3)
\end{cases}
\quad (m_1 = 0.1\ {\rm eV}),
\nonumber \\
\label{Eq:constraint_j}
\end{eqnarray}
for $0.01\le \braket{\pi_0^{(\ell^\prime\ell)}}/f_{\ell^\prime\ell}\le 1$ by the relation of 
\begin{eqnarray}
j_{\ell^\prime \ell}=  \log_{q_{\ell^\prime\ell}} \left(\frac{v}{\sqrt{2}}  \frac{\braket{\pi_0^{(\ell^\prime\ell)}}}{f_{\ell^\prime\ell}}\frac{1}{\left\vert M_{\ell^\prime \ell}\right\vert} \right),
\end{eqnarray}
and Eq.(\ref{eq:MllMaxMin}).

Hereafter, the mathematical capability of generating correct flavor neutrino mass matrix will be discussed for some selected cases. 

\begin{description}
\item[ (I) Single parameter:] 
First, we assume that the flavor structure is controlled by a single parameter of the model, e.g., $q$, $j$, $\pi_0$ or $f$ controls solely the flavor structure. In this case, there are only four possible combinations of the flavor indices:
\begin{eqnarray}
\left\vert M_{\ell^\prime \ell} \right\vert \propto
\begin{cases}
\frac{\braket{\pi_0}}{f}\frac{1}{q_{\ell^\prime\ell}^j}  &  (q_{\ell^\prime\ell}), \\
   \frac{\braket{\pi_0}}{f}\frac{1}{q^{j_{\ell^\prime\ell}}} &  (j_{\ell^\prime\ell}), \\
\frac{\braket{\pi_0^{(\ell^\prime\ell)}}}{f}\frac{1}{q^j} &  (\pi_0^{(\ell^\prime\ell)}), \\
\frac{\braket{\pi_0}}{f_{\ell^\prime\ell}} \frac{1}{q^j} &  (f_{\ell^\prime\ell}). \\
\end{cases} \nonumber
 \end{eqnarray}
These are mathematically, and probably physically, most simple assignments in this paper. We see that we can not obtain the correct flavor neutrino mass matrix in the first two cases (the single flavored $q_{\ell^\prime\ell}$ case and the single flavored $j_{\ell^\prime\ell}$ case). On the contrary, the correct flavor neutrino mass matrix can be realized in last two cases (the single flavored $\pi_0^{(\ell^\prime\ell)}$ case and the single flavored $f_{\ell^\prime\ell}$ case).

\item[ (II) Double parameters: ] 
Next, we assume that the flavor structure is controlled by double parameters of the model. Because the single flavored $q_{\ell^\prime\ell}$ or the single flavored $j_{\ell^\prime\ell}$ is incapable of generating correct flavor neutrino mass matrix, we study the effects of the collaboration between these two parameters;
\begin{eqnarray}
\left\vert M_{\ell^\prime \ell} \right\vert \propto
\frac{\braket{\pi_0}}{f}\frac{1}{q_{\ell^\prime\ell}^{j_{\ell^\prime\ell}}} &  (q_{\ell^\prime\ell} \ {\rm and} \ j_{\ell^\prime\ell}).
 \end{eqnarray}
We show that we can not obtain the correct flavor neutrino mass matrix in this case. Moreover, since the single flavored $\pi_0^{(\ell^\prime\ell)}$ and the single flavored $f_{\ell^\prime\ell}$ are capable of producing correct flavor neutrino mass matrix, we see that whether $\pi_0^{(\ell)}$ can assist the $q_{\ell^\prime\ell}$ or $j_{\ell^\prime\ell}$ to realize the correct flavor neutrino mass matrix or not. It will be shown that the following two cases are incapable of generating correct flavor neutrino mass matrix.
\begin{eqnarray}
\left\vert M_{\ell^\prime \ell} \right\vert \propto
\begin{cases}
\frac{\braket{\pi_0^{(\ell^\prime)}}}{f}\frac{1}{q_{\ell^\prime\ell}^j}  &  (\pi_0^{(\ell^\prime)}  \ {\rm and} \ q_{\ell^\prime\ell}), \\
 \frac{\braket{\pi_0^{(\ell^\prime)}}}{f}\frac{1}{q^{j_{\ell^\prime\ell}}}  &  (\pi_0^{(\ell^\prime)}  \ {\rm and} \ j_{\ell^\prime\ell}). \\
\end{cases} \nonumber
 \end{eqnarray}
We see that the other some cases, such as, 
\begin{eqnarray}
\left\vert M_{\ell^\prime \ell} \right\vert \propto
\frac{\braket{\pi_0}}{f_{\ell^\prime}}\frac{1}{q_{\ell^\prime\ell}^j}  &  (f_{\ell^\prime}  \ {\rm and} \ q_{\ell^\prime\ell}),\nonumber
 \end{eqnarray}
are also incapable of generating correct flavor neutrino mass matrix.

\item[(III) Triple parameters:] 
Finally, we assume that the flavor structure is controlled by triple parameters of the model. As an example of the triple parameters case, we see the following flavor neutrino masses 
\begin{eqnarray}
\left\vert M_{\ell^\prime \ell} \right\vert \propto
\frac{\braket{\pi_0^{(\ell^\prime)}}}{f}\frac{1}{q_{\ell^\prime\ell}^{j_{\ell^\prime\ell}}} & (\pi_0^{(\ell^\prime)}, \ q_{\ell^\prime\ell} \ {\rm and} \ j_{\ell^\prime\ell}) ,
\nonumber 
\end{eqnarray}
can be consistent with observations.
\end{description}

The detailed discussion about the capability of generating correct flavor neutrino mass matrix in these selected cases are as follows.

\

\textbf{(I-1) $q_{\ell^\prime\ell}$ dominant:} 
If the flavor structure is controlled by $q_{\ell^\prime\ell}$, the elements of the flavor neutrino mass matrix become
\begin{eqnarray}
\left\vert M_{\ell^\prime \ell} \right\vert = \frac{v}{\sqrt{2}}  \frac{\braket{\pi_0}}{f}\frac{1}{q_{\ell^\prime\ell}^j}.
\end{eqnarray}
To reproduce the flavor structure of the neutrino sector (the nine elements of the flavor neutrino mass matrix; $M_{ee},M_{e\mu}, \cdots, M_{\tau\tau}$), at least nine different values of $\left\vert M_{\ell^\prime \ell} \right\vert$ should be predicted for the fixed $\braket{\pi_0}$, $f$ and $j$; however, only two different discrete numbers 
\begin{eqnarray}
\left\vert M_{\ell^\prime \ell} \right\vert = \frac{v}{\sqrt{2}}  \frac{\braket{\pi_0}}{f} \times \underbrace{\left\{ \frac{1}{2^j}, \frac{1}{3^j} \right\}}_{2 \ {\rm numbers}},
\end{eqnarray}
could be predicted with the requirement of $q_{\ell^\prime\ell}=2,3$. We conclude that the $q_{\ell^\prime\ell}$ dominant case is excluded from observations. Thus the correct flavor neutrino mass matrix can not be realized in the $q_{\ell^\prime\ell}$ dominant case.

If we relax the requirements of $q_{\ell^\prime\ell} \in \mathbb{N}$ and allow the real and positive $q_{\ell^\prime\ell}$, there are many solutions which are consistent with observations. For example
\begin{eqnarray}
\left( 
\begin{array}{ccc}
q_{ee} & q_{e\mu} & q_{e\tau} \\
q_{\mu e} & q_{\mu\mu} & q_{\mu\tau}\\
q_{\tau e} & q_{\tau\mu} & q_{\tau\tau}
\end{array}
\right)
= \left( 
\begin{array}{ccc}
2.964 & 3.014 & 3.167 \\
3.036 & 3.029 & 2.963\\
3.077 & 2.988 & 2.977
\end{array}
\right),
\end{eqnarray}
with $\braket{\pi_0}/f=0.1$ and $j=24$ yield the flavor neutrino mass matrix in Eq.(\ref{Eq:Mbestfit}) which is consistent with the best-fit values of neutrino oscillation parameters for $m_1=0.1$ eV. 

\

\textbf{(I-2) $j_{\ell^\prime\ell}$ dominant:} 
If the flavor structure is controlled by $j_{\ell^\prime\ell}$, the elements of the flavor neutrino mass matrix become
\begin{eqnarray}
\left\vert M_{\ell^\prime \ell} \right\vert = \frac{v}{\sqrt{2}}  \frac{\braket{\pi_0}}{f}\frac{1}{q^{j_{\ell^\prime\ell}}}.
\end{eqnarray}
As same as $q_{\ell^\prime\ell}$ dominant case, at least nine different values of $\left\vert M_{\ell^\prime \ell} \right\vert$ should be predicted for the fixed $\braket{\pi_0}/f$ and $q$. Although the fourteen discrete values  
\begin{eqnarray}
\left\vert M_{\ell^\prime \ell} \right\vert = \frac{v}{\sqrt{2}}  \frac{\braket{\pi_0}}{f} \times \underbrace{\left\{ \frac{1}{2^{36}},\frac{1}{2^{37}},\cdots,\frac{1}{2^{49}} \right\}}_{14 \ {\rm numbers}},
\end{eqnarray}
could be predicted for $m_1=0.001$ eV and $q=2$ (see Eq.(\ref{Eq:constraint_j})), these values are inconsistent with observation by the following reason. 

\begin{figure}[t]
\begin{center}
\includegraphics{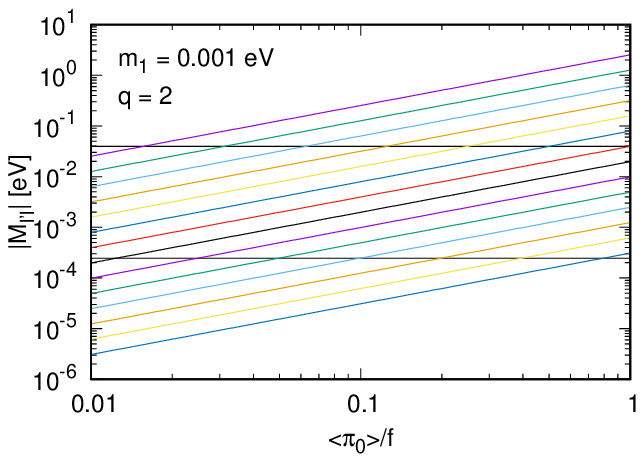} \\
\includegraphics{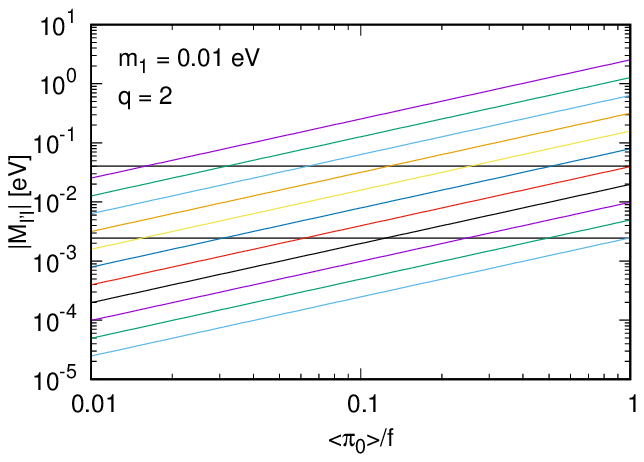} 
\includegraphics{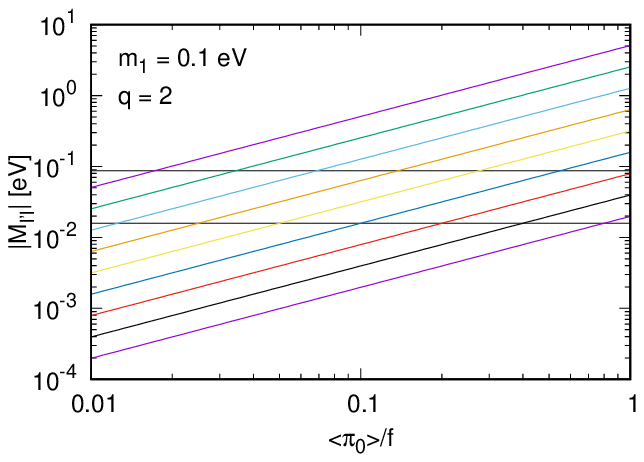}
\caption{$\left\vert M_{\ell^\prime \ell} \right\vert$ v.s. $\braket{\pi_0}/f$ for $q=2$ in the $j_{\ell^\prime\ell}$ dominant case. In the upper panel, the fourteen lines corresponding to $q^{j_{\ell^\prime\ell}}=2^{36}$ (the upper line), $q^{j_{\ell^\prime\ell}}=2^{37}$ (next to upper line) as well as to $q^{j_{\ell^\prime\ell}}=2^{49}$ (the lower line) are shown for $m_1=0.001$ eV and $q=2$. The horizontal lines show the observed upper and lower bounds of the flavor neutrino masses in the $3 \sigma$ region. The predicted nine different neutrino masses should be in this $3 \sigma$ band; however, there are maximally eight different discrete values of $\left\vert M_{\ell^\prime \ell} \right\vert$ within the $3 \sigma$ band for fixed $\braket{\pi_0}/f$. The lower panels are similar as the upper panel but for $m_1=0.01$ eV and $m_1=0.1$ eV, respectively.}
\label{fig:I-2}
\end{center}
\end{figure}

Figure \ref{fig:I-2} shows the $\left\vert M_{\ell^\prime \ell} \right\vert$ v.s. $\braket{\pi_0}/f$ in the $j_{\ell^\prime\ell}$ dominant case. In the upper panel, the fourteen lines corresponding to $q^{j_{\ell^\prime\ell}}=2^{36}$ (the upper line), $q^{j_{\ell^\prime\ell}}=2^{37}$ (next to upper line) as well as to $q^{j_{\ell^\prime\ell}}=2^{49}$ (the lower line) are shown for $m_1=0.001$ eV and $q=2$. The horizontal lines show the observed upper and lower bounds of the flavor neutrino masses  for $m_1=0.001$ eV in the $3 \sigma$ region. The nine different neutrino masses $\left\vert M_{ee} \right\vert$, $\left\vert M_{e\mu} \right\vert$, $\cdots,$ $\left\vert M_{\tau\tau} \right\vert$ should be in this $3 \sigma$ band; however, there are maximally eight different values of $\left\vert M_{\ell^\prime \ell} \right\vert$ within the $3 \sigma$ band for the fixed $\braket{\pi_0}/f$. For example, we obtain only eight numbers
\begin{eqnarray}
\left\vert M_{\ell^\prime \ell} \right\vert &=& \left\{0.000247, \  0.000495,\ 0.000990, \  0.00198, \right.\nonumber \\
&&\underbrace{\left. 0.00396,\  0.00792, \  0.0158, \ 0.0317 \right\} {\rm eV},\quad}_{8 \ {\rm numbers}} 
\end{eqnarray}
for $\braket{\pi_0}/f=0.1$. In the case of $m_1=0.01$ eV (see the lower-left panel in Fig.\ref{fig:I-2}), we have eleven different discrete values for $q=2$; however, there are maximally four different values of $\left\vert M_{\ell^\prime \ell} \right\vert$ within the $3 \sigma$ band for the fixed $\braket{\pi_0}/f$. In the case of $m_1=0.1$ eV (see the lower-right panel in Fig.\ref{fig:I-2}), we have just nine different discrete values for $q=2$; however, there are maximally three different values of $\left\vert M_{\ell^\prime \ell} \right\vert$ within the $3 \sigma$ band.  From the similar discussions, it turned out that the predicted flavor neutrino masses for $q=3$ are also inconsistent with observations. We conclude that the $j_{\ell^\prime\ell}$ dominant case for $0.001 \ {\rm eV}\le m_1 \le 0.1\ {\rm eV}$ and $0.01\le \braket{\pi_0}/f \le 1$ is excluded from the $3 \sigma$ region of the neutrino experiments. 

If we relax the requirements of $j_{\ell^\prime\ell} \in \mathbb{N}$ and allow the real and positive $j_{\ell^\prime\ell}$, there are many solutions which are consistent with observations. For example
\begin{eqnarray}
\left( 
\begin{array}{ccc}
j_{ee} & j_{e\mu} & j_{e\tau} \\
j_{\mu e} & j_{\mu\mu} & j_{\mu\tau}\\
j_{\tau e} & j_{\tau\mu} & j_{\tau\tau}
\end{array}
\right)
= \left( 
\begin{array}{ccc}
23.74 & 24.10 & 25.19 \\
24.26 & 24.21 & 23.73\\
24.55 & 23.91 & 23.84
\end{array}
\right),
\end{eqnarray}
with $\braket{\pi_0}/f=0.1$ and $q=3$ yield the flavor neutrino mass matrix in Eq.(\ref{Eq:Mbestfit}) which is consistent with the best-fit values of neutrino oscillation parameters for $m_1=0.1$ eV. 

\

\textbf{(I-3) $\pi_0^{(\ell^\prime\ell)}$ dominant:} 
\begin{figure}[t]
\begin{center}
\includegraphics{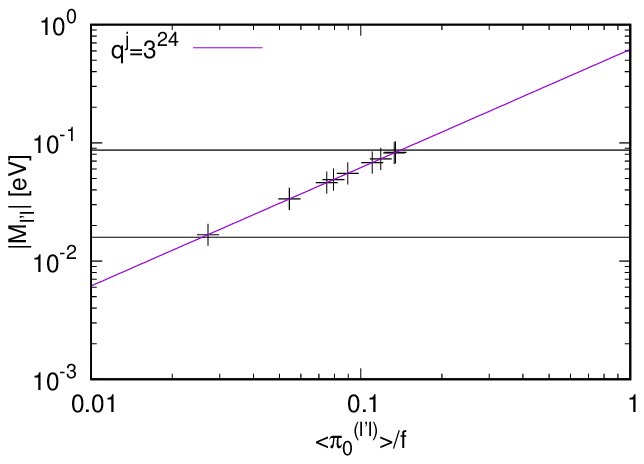}
\caption{$\left\vert M_{\ell^\prime \ell} \right\vert$ v.s. $\braket{\pi_0^{(\ell^\prime\ell)}}/f$ for $q^j=3^{24}$ in the $\pi_0$ dominated case. The horizontal lines show the observed upper and lower bounds of the flavor neutrino masses in the $3 \sigma$ region. The nine plus symbols correspond to the nine elements in Eq.(\ref{Eq:pi_I-3}). The predicted nine different neutrino masses are consistent with observations. 
}
\label{fig:I-3}
\end{center}
\end{figure}
If the flavor structure is controlled by $\braket{\pi_0^{(\ell^\prime\ell)}}$,  the elements of the neutrino mass matrix become
\begin{eqnarray}
\left\vert M_{\ell^\prime \ell} \right\vert = \frac{v}{\sqrt{2}} \frac{\braket{\pi_0^{({\ell^\prime\ell})}}}{f} \frac{1}{q^j}.
\label{Eq:M_I-3}
\end{eqnarray}
In this case, we can obtain the flavor neutrino mass matrices which are consistent with observations. For example
\begin{eqnarray}
  \frac{1}{f} \left( 
\begin{array}{ccc}
 \braket{\pi_0^{(ee)}}& \braket{\pi_0^{(e\mu)}} &  \braket{\pi_0^{(e\tau)}}\\
 \braket{\pi_0^{(\mu e)}} & \braket{\pi_0^{(\mu \mu)}} & \braket{\pi_0^{(\mu\tau)}} \\
 \braket{\pi_0^{(\tau e)}} & \braket{\pi_0^{(\tau \mu)}}& \braket{\pi_0^{(\tau\tau)}}
\end{array}
\right) 
 =
\left( 
\begin{array}{ccc}
 0.1333 & 0.08960 &  0.02715 \\
 0.07483 & 0.07929 & 0.1347 \\
 0.05440 & 0.1104 & 0.1187
\end{array}
\right),
\label{Eq:pi_I-3}
\end{eqnarray}
with $q=3$ and $j=24$ yield the flavor neutrino mass matrix in Eq.(\ref{Eq:Mbestfit}). 

Figure \ref{fig:I-3} shows the $\left\vert M_{\ell^\prime \ell} \right\vert$ v.s. $\braket{\pi_0^{(\ell^\prime\ell)}}/f$ for $q^j=3^{24}$ in the $\braket{\pi_0^{(\ell^\prime\ell)}}$ dominated case. The horizontal lines show the observed upper and lower bounds of the flavor neutrino masses in the $3 \sigma$ region. The nine plus symbols correspond to the nine elements in Eq.(\ref{Eq:pi_I-3}). We see that the predicted nine different neutrino masses are consistent with observations. 

\

\textbf{(I-4) $f_{\ell^\prime\ell}$ dominant:} 
If the flavor structure is controlled by $f_{\ell^\prime\ell}$,  the elements of the neutrino mass matrix become
\begin{eqnarray}
\left\vert M_{\ell^\prime \ell} \right\vert = \frac{v}{\sqrt{2}} \frac{\braket{\pi_0}}{f_{\ell^\prime\ell}} \frac{1}{q^j}.
\end{eqnarray}
In this case, we have the same conclusion in the $\pi_0^{(\ell^\prime\ell)}$ dominant case with the replacement
\begin{eqnarray}
\frac{\braket{\pi_0^{(\ell^\prime\ell)}}}{f}  \rightarrow \frac{\braket{\pi_0}}{f_{\ell^\prime\ell}},
\end{eqnarray}
in Eq.(\ref{Eq:M_I-3}). Thus the correct flavor neutrino mass matrix can be realized in the $f_{\ell^\prime\ell}$ dominant case.

\

\textbf{(II-1) $q_{\ell^\prime\ell}$ and $j_{\ell^\prime\ell}$ dominant:} 
If the flavor structure is controlled by $q_{\ell^\prime\ell}$ and $j_{\ell^\prime\ell}$,  the elements of the neutrino mass matrix become
\begin{eqnarray}
\left\vert M_{\ell^\prime \ell} \right\vert = \frac{v}{\sqrt{2}}  \frac{\braket{\pi_0}}{f}\frac{1}{q_{\ell^\prime\ell}^{j_{\ell^\prime\ell}}},
\end{eqnarray}
As same as $q_{\ell^\prime\ell}$ dominant case, at least nine different values of $\left\vert M_{\ell^\prime \ell} \right\vert$ should be predicted for the fixed $\braket{\pi_0}/f$. Although, fourteen discrete values for $q=2$ and nine discrete values for $q=3$ could be predicted for $m_1=0.001$ eV, it turned out that these predicted values of $\left\vert M_{ee} \right\vert$, $\left\vert M_{e\mu} \right\vert$, $\cdots,$ $\left\vert M_{\tau\tau} \right\vert$ are inconsistent with Eq.(\ref{Eq:3sigma_0.001}) for $\braket{\pi_0}/f=0.01 - 1$. We have had similar results for $m_1 = 0.001 - 0.1$ eV. We conclude that the $q_{\ell^\prime\ell}$ and $j_{\ell^\prime\ell}$ dominant case  for $0.001 \ {\rm eV}\le m_1 \le 0.1\ {\rm eV}$ and $0.01\le \braket{\pi_0}/f \le 1$ is excluded from the $3 \sigma$ region of the neutrino experiments. 

\

\textbf{(II-2) $\pi_0^{(\ell^\prime)}$ and $q_{\ell^\prime\ell}$ dominant case:} 
If the flavor structure is controlled by $\pi_0^{(\ell^\prime)}$ and $q_{\ell^\prime\ell}$,  the elements of the neutrino mass matrix become
\begin{eqnarray}
\left\vert M_{\ell^\prime \ell} \right\vert = \frac{v}{\sqrt{2}}  \frac{\braket{\pi_0^{(\ell^\prime)}}}{f}\frac{1}{q_{\ell^\prime\ell}^j}.
\end{eqnarray}
To reproduce the three elements of the flavor neutrino mass matrix $M_{\ell^\prime e}$, $M_{\ell^\prime \mu}$ and $M_{\ell^\prime \tau}$, at least three different values of $\left\vert M_{\ell^\prime \ell} \right\vert$ should be obtained for the fixed $f$ and $j$; however, only two different discrete numbers 
\begin{eqnarray}
\left\vert M_{\ell^\prime \ell} \right\vert = \frac{v}{\sqrt{2}}  \frac{\braket{\pi_0^{(\ell^\prime)}}}{f} \times \underbrace{\left\{ \frac{1}{2^j}, \frac{1}{3^j} \right\}}_{2 \ {\rm numbers}},
\end{eqnarray}
are obtained with the requirement of $q_{\ell^\prime\ell}=2,3$. We conclude that the $\pi_0^{(\ell^\prime)}$ and $q_{\ell^\prime\ell}$ dominant case is excluded from observations. 

From similar discussions, it turned out that the following assignment of the flavor indices to the flavor neutrino masses
\begin{eqnarray}
\left\vert M_{\ell^\prime \ell} \right\vert = \frac{v}{\sqrt{2}}  \frac{\braket{\pi_0^{(\ell)}}}{f}\frac{1}{q_{\ell^\prime\ell}^j},
\end{eqnarray}
as well as
\begin{eqnarray}
\left\vert M_{\ell^\prime \ell} \right\vert = \frac{v}{\sqrt{2}} \frac{\braket{\pi_0}}{f_{\ell^\prime}}\frac{1}{q_{\ell^\prime\ell}^j}, \quad
\left\vert M_{\ell^\prime \ell} \right\vert = \frac{v}{\sqrt{2}} \frac{\braket{\pi_0}}{f_\ell}\frac{1}{q_{\ell^\prime\ell}^j}, 
\end{eqnarray}
can not yield the correct flavor neutrino mass matrix.

\

\textbf{(II-3) $\pi_0^{(\ell^\prime)}$ and $j_{\ell^\prime\ell}$ dominant case:} 
If the flavor structure is controlled by $\pi_0^{(\ell^\prime)}$ and $j_{\ell^\prime\ell}$, the elements of the neutrino mass matrix become
\begin{eqnarray}
\left\vert M_{\ell^\prime \ell} \right\vert = \frac{v}{\sqrt{2}}  \frac{\braket{\pi_0^{(\ell^\prime)}}}{f}\frac{1}{q^{j_{\ell^\prime\ell}}}.
\end{eqnarray}
From the similar discussions in $q_{\ell^\prime\ell}$ and $j_{\ell^\prime\ell}$ dominant case, we conclude that the $\pi_0^{(\ell^\prime)}$ and $j_{\ell^\prime\ell}$ dominant case for $0.001 \ {\rm eV}\le m_1 \le 0.1\ {\rm eV}$ and $0.01\le \braket{\pi_0}/f \le 1$ is excluded from the $3 \sigma$ region of the neutrino experiments. 

From similar discussions, it turned out that the flavor neutrino masses  
\begin{eqnarray}
\left\vert M_{\ell^\prime \ell} \right\vert = \frac{v}{\sqrt{2}}  \frac{\braket{\pi_0^{(\ell)}}}{f}\frac{1}{q^{j_{\ell^\prime\ell}}},
\end{eqnarray}
as well as
\begin{eqnarray}
\left\vert M_{\ell^\prime \ell} \right\vert = \frac{v}{\sqrt{2}} \frac{\braket{\pi_0}}{f_{\ell^\prime}}\frac{1}{q^{j_{\ell^\prime\ell}}}, \quad
\left\vert M_{\ell^\prime \ell} \right\vert = \frac{v}{\sqrt{2}} \frac{\braket{\pi_0}}{f_\ell}\frac{1}{q^{j_{\ell^\prime\ell}}}, \end{eqnarray}
are also inconsistent with observations.

\

\textbf{(III) $\pi_0^{(\ell^\prime)}$, $q_{\ell^\prime\ell}$ and $j_{\ell^\prime\ell}$ dominant:} 
If the flavor structure is controlled by $\pi_0^{(\ell^\prime)}$, $q_{\ell^\prime\ell}$ and $j_{\ell^\prime\ell}$, the elements of the neutrino mass matrix become
\begin{eqnarray}
\left\vert M_{\ell^\prime \ell} \right\vert = \frac{v}{\sqrt{2}}  \frac{\braket{\pi_0^{(\ell^\prime)}}}{f}\frac{1}{q_{\ell^\prime\ell}^{j_{\ell^\prime\ell}}}.
\end{eqnarray}
In this case, we can obtain the flavor neutrino mass matrices which are consistent with observations. For example
\begin{eqnarray}
\frac{1}{f}\left( 
\begin{array}{ccc}
\frac{\braket{\pi_0^{(e)}}}{q_{ee}^{j_{ee}}} & \frac{\braket{\pi_0^{(e)}}}{q_{e\mu}^{j_{e\mu}}} & \frac{\braket{\pi_0^{(e)}}}{q_{e\tau}^{j_{e\tau}}} \\
\frac{\braket{\pi_0^{(\mu)}}}{q_{\mu e}^{j_{\mu e}}} & \frac{\braket{\pi_0^{(\mu)}}}{q_{\mu\mu}^{j_{\mu\mu}}} & \frac{\braket{\pi_0^{(\mu)}}}{q_{\mu\tau}^{j_{\mu\tau}}} \\
\frac{\braket{\pi_0^{(\tau)}}}{q_{\tau e}^{j_{\tau e}}} & \frac{\braket{\pi_0^{(\tau)}}}{q_{\tau\mu}^{j_{\tau\mu}}} & \frac{\braket{\pi_0^{(\tau)}}}{q_{\tau\tau}^{j_{\tau\tau}}}
\end{array}
\right)
= \left( 
\begin{array}{ccc}
\frac{0.35}{3^{27}} & \frac{0.35}{2^{43}} & \frac{0.35}{3^{27}} \\ \\
\frac{0.25}{2^{43}} & \frac{0.25}{3^{27}} & \frac{0.25}{2^{40}} \\ \\
\frac{0.4}{3^{28}} & \frac{0.4}{2^{43}} & \frac{0.4}{2^{41}}
\end{array}
\right),
\nonumber \\
\end{eqnarray}
yields the following magnitude of the flavor neutrino mass matrix
\begin{eqnarray}
\left( 
\begin{array}{ccc}
\left\vert M_{ee} \right\vert & \left\vert M_{e\mu} \right\vert & \left\vert M_{e\tau} \right\vert \\
\left\vert M_{\mu e} \right\vert & \left\vert M_{\mu\mu} \right\vert & \left\vert M_{\mu\tau} \right\vert \\
\left\vert M_{\tau e} \right\vert & \left\vert M_{\tau\mu} \right\vert & \left\vert M_{\tau\tau} \right\vert
\end{array}
\right) 
= \left( 
\begin{array}{ccc}
0.00799 & 0.00693 & 0.00780 \\
0.00495 & 0.00571 & 0.0396 \\
0.00304 & 0.00792 & 0.0317 \\
\end{array}
\right) \ {\rm eV},
\end{eqnarray}
which is consistent with observations in the $3 \sigma$ region (see Eq.(\ref{Eq:3sigma_0.01})).

 Moreover, the flavor neutrino masses
\begin{eqnarray}
\left\vert M_{\ell^\prime \ell} \right\vert = \frac{v}{\sqrt{2}}  \frac{\braket{\pi_0^{(\ell)}}}{f}\frac{1}{q_{\ell^\prime\ell}^{j_{\ell^\prime\ell}}},
\end{eqnarray}
are also consistent with observations. For example
\begin{eqnarray}
\frac{1}{f}\left( 
\begin{array}{ccc}
\frac{\braket{\pi_0^{(e)}}}{q_{ee}^{j_{ee}}} & \frac{\braket{\pi_0^{(\mu)}}}{q_{e\mu}^{j_{e\mu}}} & \frac{\braket{\pi_0^{(\tau)}}}{q_{e\tau}^{j_{e\tau}}} \\
\frac{\braket{\pi_0^{(e)}}}{q_{\mu e}^{j_{\mu e}}} & \frac{\braket{\pi_0^{(\mu)}}}{q_{\mu\mu}^{j_{\mu\mu}}} & \frac{\braket{\pi_0^{(\tau)}}}{q_{\mu\tau}^{j_{\mu\tau}}} \\
\frac{\braket{\pi_0^{(e)}}}{q_{\tau e}^{j_{\tau e}}} & \frac{\braket{\pi_0^{(\mu)}}}{q_{\tau\mu}^{j_{\tau\mu}}} & \frac{\braket{\pi_0^{(\tau)}}}{q_{\tau\tau}^{j_{\tau\tau}}}
\end{array}
\right)
= \left( 
\begin{array}{ccc}
\frac{0.12}{3^{26}} & \frac{0.1}{3^{26}} & \frac{0.11}{3^{26}} \\ \\
\frac{0.12}{2^{42}} & \frac{0.1}{2^{41}} & \frac{0.11}{2^{39}} \\ \\
\frac{0.12}{3^{27}} & \frac{0.1}{2^{41}} & \frac{0.11}{2^{39}}
\end{array}
\right),
\end{eqnarray}
yields
\begin{eqnarray}
\left( 
\begin{array}{ccc}
\left\vert M_{ee} \right\vert & \left\vert M_{e\mu} \right\vert & \left\vert M_{e\tau} \right\vert \\
\left\vert M_{\mu e} \right\vert & \left\vert M_{\mu\mu} \right\vert & \left\vert M_{\mu\tau} \right\vert \\
\left\vert M_{\tau e} \right\vert & \left\vert M_{\tau\mu} \right\vert & \left\vert M_{\tau\tau} \right\vert
\end{array}
\right) 
 = \left( 
\begin{array}{ccc}
0.00822 & 0.00685 & 0.00753 \\
0.00475 & 0.00792 & 0.0348 \\
0.00274 & 0.00792 & 0.0348 \\
\end{array}
\right)  \ {\rm eV},
\end{eqnarray}
which is consistent with  Eq.(\ref{Eq:3sigma_0.01}).

\

\section{Summary\label{section:summary}}
 The clockwork mechanism provides a natural way to obtain the hierarchical masses and couplings in a theory. In the previous studies, there are fermion clockwork models for the neutrino mixings; however, there is no scalar clockwork model for the neutrino mixings. In this paper, towards a construction of the scalar clockwork models including neutrino mixings, we have studied the mathematical capability of generating correct flavor neutrino mass matrix in a scalar clockwork model.

First, we assumed that the flavor structure is controlled by the Yukawa couplings. In this case, we can  obtain the correct flavor neutrino mass matrix by appropriate Yukawa couplings $Y_{\ell^\prime\ell}$ where $\ell^\prime, \ell = e, \mu, \tau$. 

Next, we assumed that the Yukawa couplings are extremely democratic $|Y_{\ell^\prime\ell} |=1$. In this case, the clockwork part $\left( \frac{\braket{\pi_0}}{f}\frac{1}{q^j} \right)$ should have the flavor indices $\ell^\prime$ and $\ell$. We have found that if the flavor structure is controlled by single flavored parameter $\braket{\pi_0^{(\ell^\prime\ell)}}$ or $f_{\ell^\prime\ell}$ in a scalar clockwork model, there is the mathematical capability of generating the correct flavor mass matrix in the model. In addition, if the flavor structure is controlled by triple flavored parameters $\pi_0^{(\ell^\prime)}$, $q_{\ell^\prime\ell}$ and $j_{\ell^\prime\ell}$ in a scalar clockwork model, the predicted flavor neutrino mass matrix can be consistent with observation in the $3\sigma$ region. 

Although, we have reached our main goal of our discussions to see the mathematical capability of generating correct flavor neutrino mass matrix in a scalar clockwork model, an additional discussion to see the physical availability of the model building may be required to confirm results of our discussion. Hereafter, we will show three toy models for neutrino mixings in the scalar clockwork schemes (we would like to discuss the details of the model building and phenomenological consequences such as collider experiments as a separate work in the future).

\textbf{(1) $\pi_0^{(\ell^\prime\ell)}$ dominant case:} 
First, we show a toy model for the $\pi_0^{(\ell^\prime\ell)}$ dominant case (see (I-3) in section.\ref{section:mass_matrix_cw_dominant}). In this case, to realize the $ee$-element of the flavor neutrino mass matrix 
\begin{eqnarray}
\left\vert M_{ee} \right\vert = \frac{v}{\sqrt{2}} \frac{\braket{\pi_0^{(ee)}}}{f} \frac{1}{q^j}, 
\end{eqnarray}
a clockwork chain
\begin{eqnarray}
&&\braket{\pi_0^{(ee)}}- \frac{\braket{\pi_0^{(ee)}}}{q}- \cdots -  \frac{\braket{\pi_0^{(ee)}}}{q^j}   - \cdots -   \frac{\braket{\pi_0^{(ee)}}}{q^N}  ,\nonumber \\
&& \hspace{42mm} | \\
&& \hspace{41mm} \nu_{e R}^{(e)} \nonumber
\end{eqnarray}
is required. Addition to this chain, other eight chains for $\left\vert M_{e\mu} \right\vert$, $\left\vert M_{e\tau} \right\vert$,$\cdots$, $\left\vert M_{\tau\tau} \right\vert$ are required. Therefore, a new scalar clockwork model that has nine clockwork chains in the clockwork sector is required to predict the nine flavor neutrino masses in the $\pi_0^{(\ell^\prime\ell)}$ dominant case. A candidate of the Lagrangian of this model is 
\begin{eqnarray}
\mathcal{L}_{\rm SM-CW}=\sum_{\ell^\prime, \ell} Y_{\ell^\prime\ell}  \left( \frac{\pi_j^{(\ell^\prime \ell)}}{f} \right)  \bar{\ell}^\prime_L \tilde{H} \nu^{(\ell^\prime)}_{\ell R} + {\rm h.c.}.
\end{eqnarray}
There are nine right-handed neutrinos in this model. Three of these, $\left\{ \nu_{e R}^{(e)}, \nu_{\mu R}^{(e)}, \nu_{\tau R}^{(e)}\right\}$, should interact with only left-handed electron neutrino $\nu_{e L}$ to produce $\left\vert M_{ee} \right\vert$, $\left\vert M_{e\mu} \right\vert$ and $\left\vert M_{e\tau} \right\vert$. Also, $\left\{ \nu_{e R}^{(e)}, \nu_{\mu R}^{(e)}, \nu_{\tau R}^{(e)}\right\}$ and  $\left\{ \nu_{e R}^{(e)}, \nu_{\mu R}^{(e)}, \nu_{\tau R}^{(e)}\right\}$ should interact with only $\nu_{\mu L}$ and  $\nu_{\tau L}$, respectively. These specific selection of the couplings might be realized by flavored clockwork mechanism \cite{Patel2017PRD} or the assignment of the lepton number to the clockwork sector \cite{Kitabayashi2019PRD}.  

Unfortunately, this $\pi_0^{(\ell^\prime\ell)}$ dominant model is complex. While it is found that the observed mass matrix can be reproduced, the nine clockwork chains (meaning  $\sim 100$'s of extra scalar fields) are introduced to generate the nine elements of the $3 \times 3$ mass matrix, with assumptions on their parameters and their relations (e.g., equality of some of parameters for all chains). The main cause of this complexity, nine clockwork chains, in this $\pi_0^{(\ell^\prime\ell)}$ dominant model is our assumption about the number of coupled neutrinos in a clockwork chain. In this toy model, we assume that only one neutrino flavor is permitted to couple to one clockwork chain.

A more interesting model may be achieved if different generations of neutrinos couple to different sites in a clockwork chain, which can generate hierarchies between their masses.

\textbf{(2) $\pi_0^{(\ell^\prime)}$, $q_{\ell^\prime\ell}$ and $j_{\ell^\prime\ell}$ dominant case:} 
Next, we show a toy model for the $\pi_0^{(\ell^\prime)}$, $q_{\ell^\prime\ell}$ and $j_{\ell^\prime\ell}$ dominant case (see (III) in section.\ref{section:mass_matrix_cw_dominant}). In this case, the $ee$, $e\mu$ and $e\tau$-elements of the flavor neutrino mass matrix  
\begin{eqnarray}
\left\vert M_{e\ell} \right\vert = \frac{v}{\sqrt{2}} \frac{\braket{\pi_0^{(e)}}}{f} \frac{1}{q_{e\ell}^{j_{e\ell}}}, 
\end{eqnarray}
a clockwork chain
{\small
\begin{eqnarray}
\label{Eq:summary_chane_model2}
&&\braket{\pi_0^{(e)}}-  \cdots - \frac{\braket{\pi_0^{(e)}}}{q_{ee}^{j_{ee}}}- \cdots -  \frac{\braket{\pi_0^{(e)}}}{q_{e\mu}^{j_{e\mu}}}   - \cdots -   \frac{\braket{\pi_0^{(e)}}}{q_{e\tau}^{j_{e\tau}}}  - \cdots  ,\nonumber \\
&& \hspace{22mm} |  \hspace{20mm} |  \hspace{20mm} | \\
&& \hspace{20mm} \nu_{e R}^{(e)} \hspace{16mm} \nu_{\tau R}^{(e)} \hspace{16mm} \nu_{\tau R}^{(e)} \nonumber
\end{eqnarray}
}
is required. Addition to the chain in Eq.(\ref{Eq:summary_chane_model2}), other two chains for $\left\vert M_{\mu\ell} \right\vert$ and $\left\vert M_{\tau\ell} \right\vert$ are required. Therefore, there are three clockwork chains in the clockwork sector in this model. A candidate of the Lagrangian of this model is 
\begin{eqnarray}
\mathcal{L}_{\rm SM-CW}=\sum_{\ell^\prime, \ell} Y_{\ell^\prime\ell}  \left( \frac{\pi_{j_{\ell^\prime \ell}}^{(\ell^\prime)}}{f} \right)  \bar{\ell}^\prime_L \tilde{H} \nu^{(\ell^\prime)}_{\ell R} + {\rm h.c.}.
\end{eqnarray}
The theoretical origin of inequalities in $q_{\ell^\prime \ell}$ in the same chain may be obtained by non-uniform clockwork schemes (see, for examples, Refs.\cite{Hong2019JHEP,Ben-Dayan2019PRD}).

\textbf{(3) $q_{\ell^\prime\ell}$ or $j_{\ell^\prime\ell}$ dominant case:} 
Finally,  we show a toy model for the $q_{\ell^\prime\ell}$ or $j_{\ell^\prime\ell}$ dominant cases (see (I-1) and (I-2) in section.\ref{section:mass_matrix_cw_dominant}). As we mentioned, the correct flavor neutrino mass matrix can be realized in these two cases if the requirements of $q \in \mathbb{N}$ as well as $j \in \mathbb{N}$ are relaxed in the analysis. In these cases, the correct flavor neutrino mass matrix can be realized as
\begin{eqnarray}
\left\vert M_{\ell^\prime \ell} \right\vert = \frac{v}{\sqrt{2}}  \frac{\braket{\pi_0}}{f}\frac{1}{q_{\ell^\prime\ell}^j},
\end{eqnarray}
 in the $q_{\ell^\prime\ell}$ dominant case or 
\begin{eqnarray}
\left\vert M_{\ell^\prime \ell} \right\vert = \frac{v}{\sqrt{2}}  \frac{\braket{\pi_0}}{f}\frac{1}{q^{j_{\ell^\prime\ell}}},
\end{eqnarray}
in the $j_{\ell^\prime\ell}$ dominant case with only one clockwork chain. A candidate of the Lagrangian of both models is 
\begin{eqnarray}
\mathcal{L}_{\rm SM-CW}=\sum_{\ell^\prime, \ell} Y_{\ell^\prime\ell}  \left( \frac{\pi_{j_{\ell^\prime \ell}}}{f} \right)  \bar{\ell}^\prime_L \tilde{H} \nu^{(\ell^\prime)}_{\ell R} + {\rm h.c.}.
\end{eqnarray}
A possible candidate of the theoretical origin of continuous $q$ as well as continuous $j$ may be in the continuum clockwork schemes \cite{Giudice2017JHEP,Craig2017JHEP,Giudice2018JHEP,Choi2018JHEP}. In this paper, we have discussed the capability of generating correct flavor neutrino mass matrix in the discrete scalar clockwork framework. Constructing continuum scalar clockwork model for neutrino flavor mixings with only one clockwork chain may be interesting and may be appear in future work.

Finally, we would like to comment about the issue of gauge hierarchy in the context of scalar clockwork. The electroweak scale $v$ is more than sixteen orders of magnitude smaller than the Planck scale $M_{\rm pl}$ in gravity. Within any unified theory of all interactions the small ratio $v/M_{\rm pl}$ calls for an explanation. Why the Higgs mass is so much smaller than the Planck scale. This is the gauge hierarchy problem \cite{Gildener1976PRD,Weinberg1979PLB}. (One of the solutions to the gauge hierarchy problem is realized by introducing relaxion into the theories \cite{Graham2015PRL}. The clockwork mechanisms originally introduced in the context of weak-scale relaxation \cite{Choi2016JHEP,Kaplan2016PRD}). Unlike fermionic clockwork, once a large number of scalars are utilized, each of these scalars would appear to have a hierarchy problem: why are their mass scales below the Planck scale? In the context of neutrino mass generation, the scale may be close enough to the Planck scale and the additional hierarchy problems are not severe.




%

\vspace{0.2cm}
\noindent




\end{document}